\documentclass[conference]{IEEEtran}

\usepackage{enumitem}
\usepackage{xspace}
\usepackage{amsmath}
\usepackage{caption}
\usepackage{subcaption}
\usepackage{tcolorbox}
\usepackage{hyperref}
\usepackage{tcolorbox}
\usepackage{tablefootnote}
\usepackage{color, colortbl}
\usepackage{booktabs,multicol,multirow,graphicx,tabularx,ragged2e,dcolumn}

\AtBeginDocument{%
  \providecommand\BibTeX{{%
    \normalfont B\kern-0.5em{\scshape i\kern-0.25em b}\kern-0.8em\TeX}}}

\makeatletter
\newcommand{\etal}{\emph{et al}.\@\xspace}
\newcommand*{\eg}{\emph{e.g.,}\@\xspace}
\newcommand*{\ie}{\emph{i.e.,}\@\xspace}

\newcommand{\formulas}{formul\ae\xspace}
\newcommand{\linebreakand}{%
  \end{@IEEEauthorhalign}
  \hfill\mbox{}\par
  \mbox{}\hfill\begin{@IEEEauthorhalign}
}
\makeatother

\begin{document}
\sloppy

\title{What Made This Test Flake?\\  Pinpointing Classes Responsible for Test Flakiness}
\author{\IEEEauthorblockN{Anonymous Authors}}
\author{
\IEEEauthorblockN{Sarra Habchi}
\IEEEauthorblockA{
Ubisoft \\
sarra.habchi@ubisoft.com}
\and
\IEEEauthorblockN{Guillaume Haben}
\IEEEauthorblockA{
University of Luxembourg \\
guillaume.haben@uni.lu}
\and
\IEEEauthorblockN{Jeongju Sohn}
\IEEEauthorblockA{
University of Luxembourg \\
jeongju.sohn@uni.lu}
\and
\IEEEauthorblockN{Adriano Franci}
\IEEEauthorblockA{
University of Luxembourg \\
adriano.franci@uni.lu}
\linebreakand
\IEEEauthorblockN{Mike Papadakis}
\IEEEauthorblockA{
University of Luxembourg \\
michail.papadakis@uni.lu}
\and
\IEEEauthorblockN{Maxime Cordy}
\IEEEauthorblockA{
University of Luxembourg \\
maxime.cordy@uni.lu}
\and
\IEEEauthorblockN{Yves Le Traon}
\IEEEauthorblockA{
University of Luxembourg \\
yves.letraon@uni.lu}
}

\maketitle

\begin{abstract}

Flaky tests are defined as tests that manifest non-deterministic behaviour by passing and failing intermittently for the same version of the code. 
These tests cripple continuous integration with false alerts that waste developers' time and break their trust in regression testing.
To mitigate the effects of flakiness, both researchers and industrial experts proposed strategies and tools to detect and isolate flaky tests. 
However, flaky tests are rarely fixed as developers struggle to localise and understand their causes. Additionally, developers working with large codebases often need to know the sources of non-determinism to preserve code  quality, \ie avoid introducing technical debt linked with non-deterministic behaviour, and to avoid introducing new flaky tests. To aid with these tasks, we propose re-targeting Fault Localisation techniques to the flaky component localisation problem, \ie pinpointing program classes that cause the non-deterministic behaviour of flaky tests. In particular, we employ Spectrum-Based Fault Localisation (SBFL), a coverage-based fault localisation technique commonly adopted for its simplicity and effectiveness. We also utilise other data sources, such as change history and static code metrics, to further improve the localisation. 
Our results show that augmenting SBFL with change and code metrics ranks flaky classes in the top-1 and top-5 suggestions, in 26\% and 47\% of the cases. 
Overall, we successfully reduced the average number of classes inspected to locate the first flaky class to 19\% of the total number of classes covered by flaky tests. 
Our results also show that localisation methods are effective in major flakiness categories, such as concurrency and asynchronous waits, indicating their general ability to identify flaky components. 

\end{abstract}


\maketitle

\section{Introduction}
Regression testing is a key component of continuous integration (CI) that checks whether code changes integrate well in the codebase without breaking any existing functionality. To this end, it is assumed that failing tests indicate the presence of faults, introduced by the latest changes. However, some tests break this assumption by failing for reasons other than faults, as for instance,  they exhibit non-deterministic behaviour, thereby sending confusing signals to developers. Such tests are usually called \textit{flaky tests}.

Academic and industrial reports have emphasised the adverse effects of test flakiness in software development. Specifically, Google reported that 16\% of their tests manifested some level of flakiness, while more than 90\% of their test transitions, either to failing or passing, were due to flakiness~\cite{LeongSPTM19}. 
As the \textit{de facto} approach for detecting flaky tests is to rerun them \cite{habchi2021qualitative, GruberFraser22}, detecting large numbers of flaky tests can be time- and resource-consuming. Indeed, Google reports that between 2 to 16\% of their CI resources are dedicated in rerunning flaky tests~\cite{Micco2017}. It is noted  that other companies, like Microsoft~\cite{Lam2020b},  Spotify~\cite{FlakinessSpotify} and Mozilla~\cite{Rahman2018}, also report similar issues when dealing with test flakiness. 

Perhaps more importantly, test flakiness affects team productivity and software quality \cite{habchi2021qualitative}. This is because flaky failures interrupt the CI and make developers waste time in investigating false issues~\cite{GTAC2016,Eck2019,LeongSPTM19,habchi2021qualitative}. Additionally, the accrual of flaky tests breaks the trust in regression testing, leading developers to disregard legitimate failure signals believing them to be false~\cite{habchi2021qualitative, GruberFraser22}. This situation often results in faults slipping into production systems~\cite{Rahman2018}. Moreover, code quality is often linked with the level of flakiness incurred \cite{habchi2021qualitative} and thus, developers need to know where it comes from and understand the causes of flakiness to avoid introducing and spreading it. 

Given the adverse effects of test flakiness, engineers and researchers aim at developing 
detection techniques that can predict whether a test is potentially flaky. These approaches rely on a number of runs and re-runs, such as \textsc{iDFlakies}~\cite{Lam2019iDFlakies} and \textsc{Shaker}~\cite{Silva2020}, coverage analysis like \textsc{DeFlaker}~\cite{Bell2018}, or static and dynamic test features~\cite{alshammari2021flakeflagger,Haben2021,Pinto2020,Dong2021,Camara2021VocabExtendedReplication,Camara2021a,Fatima2021}.  
Evaluated on open-source projects, these approaches showed promising detection accuracy and considerably decreased the amount of time and resources needed to detect flaky tests. 

Although flakiness detection methods are important, alone, they cannot reduce the prevalence of test flakiness. This is because on the one hand there are only partial approaches to  the problem,  such as  \textsc{iFixFlakies}~\cite{Shi2019iFix} and \textsc{Flex}~\cite{FLEX} that are only applicable to specific cases, and the inherent difficulties in isolating/controlling the flakiness causes on the other. For instance, \textsc{iFixFlakies}~\cite{Shi2019iFix} fixes order-dependent tests by identifying helper statements in other tests, whereas \textsc{Flex}~\cite{FLEX} identifies  assertion bounds that minimise flakiness stemming from algorithmic randomness. At the same time, many prevalent categories of flakiness, \eg  Asynchronous Waits and Concurrency~\cite{Gruber2021,romano2021empirical,Luo2014, Eck2019}, remain unaddressed by fixing approaches. This is mainly due to the difficulty of identifying  and controlling the cause of flakiness~\cite{Eck2019}. 

Flakiness root cause localisation is both important and difficult. It is important since it allows developers to understand the sources of flakiness, hence enabling better control of non-determinism. It is also difficult because of the difficulty to reproduce failures, the diversity in potential issues, \eg time and network, and the large scope of potential culprits, \eg the tests, the code under test (CUT), and the infrastructure~\cite{Gruber2021}. Consequently, practitioners struggle to identify the causes of non-determinism in their codebases that trigger flakiness and consider this step as the main challenge in automating flakiness mitigation strategies~\cite{Eck2019}. 

In this paper, we address this challenge by re-targeting Fault Localisation (FL) techniques in order to help identify components (program classes in particular) that are responsible for the non-deterministic behaviour of flaky tests. For the sake of simplicity, we refer to these classes as \textit{flaky classes}. Such techniques can be useful to support the analysis  of codebases and of flaky tests. Thus, given a failure, either known as flaky or unknown, engineers can rely on localisation methods to investigate the specific scenario (condition) that causes the test transition. Additionally, flakiness localisation techniques can help with code comprehension and make engineers aware of code areas linked with flaky behaviour, assisting them in both development and testing tasks. 

In view of this, we investigate the appropriateness of a variety of fault localisation methods, such as Spectrum-Based Fault Localisation (SBFL), change history metrics, and static code metrics in identifying flaky classes. 
Our study aims to answer the following four research questions:

\begin{itemize}
    \item \textbf{\textsc{RQ1:}} Are SBFL-based approaches effective in identifying flaky classes?
    \item \textbf{\textsc{RQ2:}} How do code and change metrics contribute to the identification of flaky classes?
    \item \textbf{\textsc{RQ3:}} How can ensemble learning improve the identification of flaky classes?
    \item \textbf{\textsc{RQ4:}} How does an SBFL-based approach perform for different flakiness categories?
\end{itemize}

To answer these questions, we analyse five Open Source projects where test flakiness has been fixed during the project evolution. Our analysis shows that: 
\begin{itemize}
    \item An ensemble of models based on SBFL, change, and size metrics, yields the best results, with 61\% of flaky classes in the top 10 and 26\% of them at the top. 
    This method also reduces the average effort wasted by developers to 19\% of the effort spent when inspecting all classes covered by the flaky test.
    \item The ensemble method is effective for major flakiness categories. Concurrency and Asynchronous Waits are identified effectively, with 38\% and 30\% of their flaky classes ranked at the top, respectively.
\end{itemize}
    
    
        
    
     

To facilitate the reproducibility of this study, we provide all used scripts,  the set of collected flaky classes, and detailed results in a comprehensive package\footnote{\url{https://github.com/serval-uni-lu/sherlock.replication}}.

\section{Data Collection}
\label{sec:data_collection}
The objective of our study is to assess the effectiveness of FL techniques in identifying flaky classes.
To achieve this, we need a set of flaky tests for which the responsible classes are already known. 
For this, we rely on flakiness-fixing commits as they provide information about classes that were modified as part of the fix.
Our assumption is that such classes are, at least, part of the root cause. 
To collect flaky classes, we followed a four-step process.

\paragraph{Search} 
This step aims to identify Java projects containing the highest number of flakiness-fixing commits. For this, we relied on two sets of projects to consider. 
We built the first set by using the SEART GitHub Search Engine~\cite{githubsearch}. Out of the 81,180 available Java projects, we selected the top 200 projects for each of those criteria: number of commits, contributors, stars, releases, issues, and files. This sorting was made with the aim of finding the bigger and more complex projects, thus maximising our chance to find flakiness-fixing commits. 
Keeping only unique projects in those sets, we ended up with a first list of 902 projects. As a second set, we use the 187 projects available in the \textsc{iDFlakies} dataset~\cite{Lam2019iDFlakies}. 
For each of the 1,089 projects, 902 from the first and 187 from the second set, we query the GitHub API looking for commits with messages containing the keyword \textit{flaky}. This led to the identification of 16,501 commits.
We look further into whether these commits are truly suitable for our purpose through the following processes. 

\newcommand\mch[2]{\multicolumn{1}{>{\centering\arraybackslash}b{#1}}{#2}}

\begin{table}[t]
\vspace{-0.5em}
\begin{center}
\caption{Collected Data. \textit{ffc:} number of flakiness-fixing commits. \textit{all:} number of commits in the project.
\vspace{-0.5em}
\centering}
\centering
\label{tab:SD-info}
\scalebox{0.8}{
\begin{tabular}{l|cc|cc|cc}
\toprule
\textbf{Proj. } & \multicolumn{2}{c|}{\textbf{\#Commits}} & \multicolumn{2}{c|}{\textbf{\#Tests}} & \multicolumn{2}{c}{\textbf{\#Classes}} \\
 & ffc & all &  min - max & avg & min -- max & avg \\
\midrule
Hbase & 8 & 18,990 & 138 - 2,089 & 1,113 & 734 -- 1366 & 1053.4 \\
Ignite & 14 & 27,903 & 15 - 1,018 & 174 & 72 -- 1767 & 1262.3 \\
Pulsar & 10 & 8,516 & 194 - 1,326 & 626 & 171 -- 422 & 259.7 \\
Alluxio & 3 & 32,560 & 315 - 694 & 473 & 131 -- 817 & 360.3 \\
Neo4j & 3 & 71,824 &  21 - 5,782 & 2,139 & 40 -- 1663 & 581.3 \\
\midrule
Total & 38 &  &  15 - 5,782 & 905 & 40 -- 1767 & 820.2 \\
\bottomrule
\end{tabular}
}
\end{center}
\vspace{-5mm}
\end{table}


\paragraph{Inspection}
The objective of this step is to filter commits that do not provide a clear indication about the flaky class. Hence, we look for flakiness-fixing commits containing any of the following keywords: \textit{fix, repair, solve, correct, patch, prevent}. Then, we analyse each commit and keep the ones that:
\begin{itemize}[wide=0pt,noitemsep,topsep=0pt]
    \item The fix affects the code under test 
    (not only the test itself);
    \item The changes are atomic enough (\ie containing only relevant changes) allowing us to discern the flaky class(es). 
\end{itemize}

This led to the selection of 85 commits from five projects. We further discarded 22 commits for which the flaky tests or commit were not retrievable (\eg rejected pull request), leaving 63 commits in the end.

\paragraph{Test execution}
This step aims to select commits that are usable in our evaluation.
Our first question inspects the effectiveness of SBFL, a technique that requires a coverage matrix indicating the classes covered by each test.
Hence, for a commit to be usable in our analysis, its test suite should be runnable allowing us to extract the coverage matrix.
To ensure this, we used  \textsc{GZoltar}\footnote{\url{https://github.com/GZoltar/gzoltar/blob/master/com.gzoltar.ant/}}, a Maven plugin that allows collecting coverage information for each commit.
For 11 commits, we were unable to run \textsc{GZoltar} due to an incompatible Java version. We also found that the flakiness patches were irrelevant in 10 commits.
For instance, some commits were fixing modules in other programming languages or modifying non-source code files.
Lastly, we filtered out four additional commits since the reported flaky failures were not \textit{flaky test failures}.
Consequently, 
we dropped 25 commits in addition. Table~\ref{tab:SD-info} summarises the retained projects.
The complete list of flakiness-fixing commits is available in our replication package. 

\paragraph{Extraction}
For each collected flakiness-fixing commit, we retrieve the source code, the test suite, the fixed flaky test, and the flaky class. To retrieve the flaky classes, two authors manually analysed the commit diff and message to identify them.
Overall, the identification was obvious since we selected atomic commits beforehand.
Hence, there were no disagreements between the authors at this step.
The identified classes are considered the ground truth of our study. 

\section{Study Design}





\subsection{RQ1 - Effectiveness}\label{sub:rq1_effectiveness}
\subsubsection{Motivation}\label{subsub:rq1_motivation}
The objective of our study is to investigate the usability of well-founded FL techniques to help in mitigating flaky tests.
The literature on FL proposes a wide variety of categories such as ML-based techniques~\cite{Lou:2021:fse,Li:2019:issta,briand2007}, mutation-based techniques~\cite{Papadakis:2015sf,Hong:2015db}, and qualitative reasoning-based techniques~\cite{perez2018leveraging}.
Nonetheless, spectrum-based fault localisation remains one of the most distinguished FL categories thanks to its effectiveness and simplicity~\cite{wong2016survey}. 
SBFL requires only the test coverage matrix to compute the likelihood for a code entity to include the root cause of an observed test failure. 
The main assumption of SBFL is that code entities covered by more failing tests and fewer passing tests are more suspicious than those less covered by failing tests and more by passing tests~\cite{renieres2003fault}.
This assumption can be revised to identify the root causes of flaky tests instead of bugs.
In particular, if we separate tests into two groups: \textit{flaky} and \textit{stable}, instead of \textit{failing} and \textit{passing}, we can leverage the coverage matrix to rank classes based on their correlation with flaky tests. 
In this case, the assumption would be that classes covered by more flaky tests and fewer stable tests have a higher chance to be responsible for test flakiness.
In this RQ, we assess the effectiveness of this adaptation of SBFL in identifying flaky classes.

\subsubsection{Approach}\label{subsub:rq1_approach}
Relying on the data collected in Section~\ref{sec:data_collection}, we use the \textsc{GZoltar} plugin to run the test suites of each commit and build coverage matrices.
Based on these matrices, we compute for each class the spectrum data: $(e_{s}, e_{f} , n_{s}, n_{f})$.
In our case, for each class, $e_{s}$ and $e_{f}$ represent the number of stable and flaky tests executing it, respectively.
On the other hand, $n_{s}$ and $n_{f}$ represent the number of stable and flaky tests that do not execute it, respectively.
To compute classes' suspiciousness scores, we inject these spectrum data in classical SBFL formulæ.
Table~\ref{tab:formulae} summarises the four formulæ adopted in our study with the necessary adaptations for flakiness.
For DStar, the notation ‘*’ is a variable that we set to 2 based on the recommendation of Wong \etal~\cite{wong-dstar}.
With each formula, we compute the suspiciousness scores of each class and then rank them in descending order: classes with the highest scores are ranked first.

Recently, it has been  theoretically proven that no SBFL formula can outperform all others~\cite{Yoo:2014fv}. In addition, Xuan and Monperrus proposed a new approach that learns to combine multiple SBFL \formulas~\cite{monperrus-ICSME}. Their approach, called Multric, successfully outperformed all the input \formulas, opening a trend to use multiple \formulas to overcome the limitation of using a single SBFL formula~\cite{B.-Le:2016yu,zou2019empirical,Li:2019:issta}. 
Following this trend, we used Genetic Programming to evolve a new formula that combines all four SBFL \formulas. 

Genetic Programming (GP) evolves a solution (\ie a program) for a given problem under the guidance of a (fitness) function. 
GP can also  generate non-linear models and learn a model flexibly from input instead of defining a fixed formula. Hence, GP was employed to generate risk evaluation \formulas for fault localisation~\cite{Yoo:2017ss,sohn-TSE}. 
For the same reasons, we employ GP to evolve a model (\ie a formula) for the flaky class identification problem. 
We configure the GP to have a population of 40 individuals and to stop and return the best model found so far after 100 generations. 
Each individual in the population denotes a single candidate formula and is generated using (i) six arithmetic operators (subtraction, addition, multiplication, division, square root, and negation) and (ii) the features that GP takes as input.
We define our fitness function as the average ranking of flaky classes. To make most of the data and avoid overfitting, we use ten-fold cross-validation, using one fold for test and the others for training. We also normalise all input data between 0 and 1 using min-max normalisation. 
Finally, to compensate for the inherently stochastic nature of GP, we run GP 30 times with different random seeds and report the results of a model with the median fitness. 
We used DEAP v.1.3.1~\cite{Fortin:2012aa}. 

\begin{table}
\vspace{-0.5em}
\begin{center}
\caption{SBFL formulae adapted to flakiness.
\vspace{-0.5em}
\centering}
\label{tab:formulae}
\begin{tabular}{lc} 
\toprule
 \textbf{{Name}} & \textbf{{
Formula}}   \\  \hline
Ochiai~\cite{Abreu:2006yf} & $\frac{e_f}{\sqrt{(e_f + n_f)(e_f + e_s)}}$ \\ 
Barinel~\cite{abreu2009spectrum} & $1 - \frac{e_s}{e_s + e_f}$ \\ 
Tarantula~\cite{Jones:2001vn,Jones:2002kx} & $\frac{\frac{e_f}{e_f+n_f}}{\frac{e_f}{e_f+n_f}+\frac{e_s}{e_s+n_s}}$ \\ 
DStar~\cite{wong-dstar} & $\frac{e_f^*}{e_s * n_f}$ \\\bottomrule
\end{tabular}
\end{center}
\vspace{-6mm}
\end{table}
\subsection{RQ2 - Code and change metrics}
\subsubsection{Motivation}
The objective of this question is to explore the benefits of augmenting the SBFL technique with additional signals from the software.
Recent studies showed that the performances of SBFL can be improved by incorporating signals from code and change metrics.
More specifically, Sohn and Yoo~\cite{sohn-TSE} showed that combining SBFL with code and change metrics widely adopted in the fault prediction community~\cite{McI:2018:tse}, such as age, change frequency (\ie churn), and size, can significantly improve the approach's performances.
The assumption is that code entities with higher complexity and change frequency are more likely to be faulty.
Several studies suggested that the test size and complexity can also be an indicator of flakiness~\cite{Pinto2020,King2018,Camara2021a}. %
However, it is unclear if such metrics correlate also with classes that are responsible for test flakiness.
Therefore, in this RQ, we assess the benefits of these metrics in spotting flaky classes.
Besides these metrics, we investigate the effects of metrics that are specific to the nature of flaky tests.
Multiple empirical studies analysed the root causes of flakiness and showed that the main categories are: Async Waits, Concurrency, Order-dependency, Network, Time, I/O operations, Unordered collections and Randomness~\cite{Luo2014,Parry2021,Lam2020a,Gruber2021}.
We derived a list of static metrics that describe each of these categories in Java projects.
We exclude order-dependency because order-dependent tests generally stem from tests themselves instead of the CUT, thus, they are not concerned by our approach.
In the following, we describe our approach for (i) calculating these metrics and (ii) defining a FL formulae based on them.

\subsubsection{Approach}

\paragraph{Metric collection}
Table~\ref{tab:metrics} summarises the full list of metrics used in our study.
To compute these metrics, we first retrieve the source code of the project at the commit of interest (\ie the parent commit of the flakiness-fixing commit identified by the data collection step).
Then, for calculating flakiness-specific metrics, we use Spoon~\cite{spoon}.
Spoon is a framework for Java-based program analysis and transformation that allows us to build an abstract syntax tree and a call graph.
Using the graph and tree, we extract classes and their metrics (\eg \#COPS and \#ROPS). 
For size metrics, we also use these code analysis results from Spoon (\eg DOI). 
As for change metrics, we analyse the change history and extract the following information: the date of each commit, files modified and renamed by each commit, and authors of individual commits. Using this information, we compute the three change metrics: Unique Changes, Age, and Developers.

\begin{table}
\vspace{-0.5em}
\begin{center}
\caption{Code and change metrics used to augment SBFL.\centering}
\vspace{-0.5em}
\label{tab:metrics}
\begin{tabularx}{\columnwidth}{l|l|X} 
\cmidrule[\heavyrulewidth]{2-3}
\textbf{{}} & \textbf{{Metric}} & \textbf{{Definition}} \\ \midrule
\multirow{10}{*}{\rotatebox[origin=c]{90}{Flakiness}} & \#TOPS & Number of time operations performed by the class.\\ \cline{2-3}
& \#ROPS & Number of calls to the \texttt{random()} method in the class.\\ \cline{2-3}
& \#IOPS & Number of input/output operations performed by the class.\\ \cline{2-3}
& \#UOPS & Number of operations performed on unordered collections by the class.\\ \cline{2-3}
& \#AOPS & Number of asynchronous waits in the class.\\ \cline{2-3}
& \#COPS & Number of concurrent calls in the class.\\ \cline{2-3}
& \#NOPS & Number of network calls in the class. \\
\midrule 

\multirow{3}{*}{\rotatebox[origin=c]{90}{Change}} & Changes & 
Number of unique changes made on the class. \\ \cline{2-3}
& Age & 
Time interval to the last changes made on the class. \\ \cline{2-3}
& Developers & Number of developers contributing to the class. \\ \midrule

\multirow{3}{*}{\rotatebox[origin=c]{90}{Size}} & LOC & The number of lines of code.\\ \cline{2-3}
& CC & Cyclomatic complexity.\\ \cline{2-3}
& DOI & Depth of inheritance.\\ 
\bottomrule
\end{tabularx}
\end{center}
\vspace{-4mm}
\end{table}

\paragraph{Ranking model}
Similarly to RQ1, we use GP in order to generate models that combine our metrics with suspiciousness scores generated by SBFL \formulas.
In particular, for each type of metrics (\ie flakiness, size, and change), we evolve a model that takes as input its metrics with SBFL scores and outputs a ranking for each candidate class.
Afterwards, we compare the performances of these models to infer the contribution of each type of metrics.

\subsection{RQ3 - Ensemble method}\label{sec:rq3_ensemble_method}
\subsubsection{Motivation}
This question explores the potential for improvement by exploiting all the \formulas generated using GP while at the same time making the most of the resources spent on model generation.
For this aim, we use voting as our ensemble learning method.
We opted for voting since it does not require an additional cost for model generation and its effectiveness has already been demonstrated by previous fault localisation studies~\cite{Sohn2021ea,Sohn2019aa}

\subsubsection{Approach}
Voting between models is performed in two phases: candidate selection and voting. During the candidate selection phase, all the participating models compute their own suspciousness scores for the candidates. A candidate, in our case, is an individual class of the CUT. 
Individual models compute their own suspiciousness scores for the candidates and select those placed within the top $N$ as their candidates to vote.
In the voting phase, each model votes for its own top $N$ candidates. If $M$ number of models participate in the voting, we can have the maximum $N \times M$ number of voted candidates in total. The votes from the models are then aggregated, and the voted candidates are reordered from the most voted to the least voted.

Previous studies on voting-based FL showed that varying the number of votes that each candidate receives based on its actual rank in individual models can improve the localisation performance even further~\cite{Sohn2021ea,Sohn2019aa}.
Hence, rather than assigning the same number of votes to each candidate, we allow individual models to cast a different number of votes for each candidate based on its location in the ranking. 
For instance, a candidate ranked at the top will obtain a complete one vote, whereas a candidate ranked in the third place will get $\frac{1}{3}$ vote. 
As mentioned in~\ref{Tie-breaking}, candidates can be tied with other candidates since their ranks are computed from ordinal scores. When a candidate fails to be in the top $N$ due to being tied with others, we allow every tied candidate ($c$) to receive the following number of votes:
$votes = \frac{1}{rank_{best}(c) \times n_{tied}(c)}$ votes. 
Here $rank_{best}$ denotes the best (highest) rank a tied candidate can have, and $n_{tied}$ is the total number of tied candidates, including itself. The equation below summarises the number of votes a candidate ($c$) can obtain. $rank(c)$ is the rank of the candidate $c$. 

\vspace{-4mm}
\[
\left\{ 
  \begin{array}{ c l }
    \frac{1}{rank(c)} & \quad \textrm{if } rank(c) \leq N \\
    \frac{1}{rank_{best}(c) \times n_{tied}(c)} & \quad \textrm{if } rank_{best}(c) \leq N \\
    0                 & \quad \textrm{otherwise}
  \end{array}
\right.
\]

\subsection{RQ4 - Flakiness categories}
\subsubsection{Motivation}
The literature on flaky tests reports different categories of flakiness~\cite{Luo2014,Parry2021,Lam2020a,Gruber2021}.
These categories can manifest differently both in the test and CUT and as a result the identification of flaky classes can also be affected by such differences.
That is, a technique might identify decently the classes responsible for non-deterministic network operation, but struggles in pinpointing classes causing race conditions.
This RQ aims to investigate the performances of an SBFL-based approach among distinct flakiness categories.

\subsubsection{Approach}
Many studies manually analysed flakiness-fixing commits to categorise them~\cite{Luo2014,Thorve2018} based on their commit message and code changes.
In our study, we followed a similar process where two authors manually analysed the commits separately to assign them to one of the categories derived by Luo~\etal~\cite{Luo2014}.
As our manual analysis does not intend to build a new taxonomy or identify new categories, it is reasonable to adopt an existing taxonomy as reference.
The two authors had a disagreement over one commit, where one author only suggested one category whereas the other suggested two categories.
After discussion, the authors decided to keep two categories to avoid discarding relevant information.
The results of this analysis are available in our replication package.
After labelling the flakiness-fixing commits, we analyse the performance of our SBFL-based approach among different flakiness categories.

\subsection{Evaluation metrics}
For the evaluation of our approach, we use two metrics: accuracy and wasted effort. 
Both \textit{acc@n} and \textit{wef} are based on the absolute number of code entities instead of percentages.
This conforms to the recommendations of Parnin and Orso~\cite{parnin} who suggested that absolute metrics reflect the actual amount of efforts required from developers better than percentages. 
The accuracy (\textit{acc@n}) calculates the number of cases where the flaky classes were ranked in the top $n$.
In our study, we report the \textit{acc@n} with 1, 3, 5, and 10 as n values. 
In the cases of multiple flaky classes, we consider the flaky class to be among the top $n$, if at least one of the flaky classes is.
The second metric, wasted effort (\textit{wef}), allows us to measure the effort wasted while searching for the flaky class. It is formally defined as~\cite{monperrus-ICSME}:
\[ wef = |{susp(x) > susp(x*)}| + |{susp(x) = susp(x*)}|/2 + 1/2\]
Where $susp()$ provides the suspiciousness score of the class $x$, $x*$ is the flaky class, and $|.|$ provides the number of elements in the set. Accordingly, \textit{wef} measures the absolute number of classes inspected before reaching the real flaky class $x*$. 

For our approach to be useful for developers, it should provide guidance beyond currently available information. 
When a program fails due to flaky tests, one thing that can be helpful to identify the cause is a list of classes covered by the flaky tests.
Hence, in this paper, we count the total number of classes covered by flaky tests (\ie our baseline) and compare it with the number of classes inspected to locate a flaky class (\ie \textit{wef}$+1$). 
More specifically, in addition to the two absolute metrics, we measure the relative effort defined as:
\[R_{wef} = \frac{100 \times (wef + 1)}{\text{\# of classes covered by flaky tests}}\text{, } 0 < R_{wef} \leq 100\]

If $R_{wef}$ is smaller than 50, we consider our approach to outperform the baseline since it saves more than the expected effort (\ie average) of the baseline.

\subsection{Tie-breaking}
\label{Tie-breaking}
Both SBFL and our evolved \formulas compute an ordinal score for each class. As a result, multiple classes can have the same score, being tied to each other. Ties are generally harmful as they force developers to inspect more classes. Among various tie-breakers introduced and adopted to handle this problem~\cite{xu2011ties}, we use a max tie-breaker that assigns the lowest rank (\ie the maximum) to all tied entities. We choose the max tie-breaker to avoid overinterpretation of the results. 
\section{Study Results}
\label{sec:results}

\subsection{RQ1 - Effectiveness}

Table~\ref{tab:RQ1} shows the localisation results of SBFL \formulas.
Among the four SBFL \formulas, Dstar yields the worst results both in accuracy and wasted effort, while the other three perform similarly. Out of 38 analysed flaky classes, Dstar ranks 18 (47\%) in the top 10. Ochiai, which performs the best, places 53\% of flaky classes (\ie 21) within the top 10 and 16\% (6) at the top. Nevertheless, regardless of which formula we use, our SBFL-based approach outperforms the baseline of inspecting classes covered by flaky tests: for all four SBFL \formulas, $R_{wef}$ is always smaller than 50 in total, 
especially in its median. It is worth noting that since the total number of classes covered by flaky tests differs in each flaky commit, $R_{wef}$ does not always concur with \textit{wef}. For Ochiai, $R_{wef}$ reduces to 6, meaning we only need to inspect 6\% of the classes covered by flaky tests.  

\begin{table*}[ht]
\caption{RQ1: Effectiveness of SBFL \formulas. 
(\#) denotes the total number of flaky commits for each project\centering.
The row \textit{Perc} contains the percentage of flaky commits whose triggering flaky classes are ranked in the top $n$; these values are computed only for \textit{acc@n}. 
\label{tab:RQ1}}
\centering
\scalebox{0.72}{
\begin{tabular}{c|p{0.26cm}p{0.26cm}p{0.26cm}p{0.45cm}|rr|p{0.26cm}p{0.26cm}p{0.26cm}p{0.45cm}|rr|p{0.26cm}p{0.26cm}p{0.26cm}p{0.45cm}|rr|p{0.26cm}p{0.26cm}p{0.26cm}p{0.45cm}|rr}
\toprule
& \multicolumn{6}{c|}{\textbf{Dstar}} & \multicolumn{6}{c|}{\textbf{Ochiai}} & \multicolumn{6}{c|}{\textbf{Tarantula}} & \multicolumn{6}{c}{\textbf{Barinel}} \\
\textbf{Proj. (\#)} & \multicolumn{4}{c}{acc} & \multicolumn{2}{c|}{wef (R$_{wef}$)} & \multicolumn{4}{c}{acc} & \multicolumn{2}{c|}{wef (R$_{wef}$)} & \multicolumn{4}{c}{acc} & \multicolumn{2}{c|}{wef (R$_{wef}$)} & \multicolumn{4}{c}{acc} & \multicolumn{2}{c}{wef (R$_{wef}$)} \\
& @1 & @3 & @5 & @10 & mean & med & @1 & @3 & @5 & @10 & mean & med & @1 & @3 & @5 & @10 & mean & med & @1 & @3 & @5 & @10 & mean & med \\
\midrule

Hbase (8) & 0 & 3 & 4 & 4 & 33.0 (17) & 7 (5) & 2 & 5 & 5 & 5 & 14.9 (13) & 1 (4) & 1 & 4 & 4 & 5 & 11.9 (12) & 4 (4) & 1 & 4 & 4 & 5 & 11.6 (12) & 4 (4) \\
ignite (14) & 0 & 2 & 2 & 2 & 214.7 (21) & 31 (4) & 0 & 3 & 3 & 4 & 212.0 (20) & 20 (4) & 0 & 3 & 3 & 4 & 177.1 (17) & 20 (4) & 0 & 3 & 3 & 4 & 175.1 (17) & 20 (4) \\
Pulsar (10) & 1 & 3 & 6 & 9 & 9.9 (21) & 4 (6) & 3 & 5 & 6 & 9 & 9.2 (13) & 3 (6) & 3 & 5 & 6 & 9 & 9.2 (13) & 3 (6) & 3 & 5 & 6 & 9 & 9.2 (13) & 3 (6)\\
Alluxio (3) & 0 & 0 & 0 & 1 & 60.7 (43) & 72 (31) & 0 & 0 & 0 & 1 & 71.0 (46) & 72 (41) & 0 & 0 & 0 & 0 & 92.7 (59) & 73 (58) & 0 & 0 & 0 & 0 & 105.3 (66) & 87 (65) \\
Neo4j (3) & 1 & 2 & 2 & 2 & 12.0 (41) & 1 (18) & 1 & 2 & 2 & 2 & 12.0 (41) & 1 (18) & 1 & 2 & 2 & 2 & 23.0 (43) & 1 (18) & 1 & 2 & 2 & 2 & 23.7 (43) & 1 (18) \\

\midrule
Total  (38) & 2 & 10 & 14 & 18 & 94.4 (24) & \textbf{11 (17)} & 6 & 15 & 16 & 21 & 90.2 (21) & 7 (6) & 5 & 14 & 15 & 20 & 79.3 (21) & 8 (7) & 5 & 14 & 15 & 20 & 79.6 (21) & 8 (7) \\
Perc (\%) & 5 & 26 & 37 & \textbf{47} & - & - & \textbf{16} & \textbf{39} & 42 & \textbf{55} & - & - & 13 & 37 & 39 & \textbf{53} & - & - & 13 & 37 & 39 & \textbf{53} & - & - \\
\bottomrule
\end{tabular}
}
\vspace{-4mm}
\end{table*}


Table~\ref{tab:RQ1-GP} presents the evaluation results of our GP model evolved to combine the four SBFL \formulas.
As explained in Section \ref{sub:rq1_effectiveness}, we report only the results of the model with the median fitness among 30 models. 
In contrast to what we expected from combining the four SBFL \formulas using GP, we fail to observe any meaningful improvement compared to the results of Ochiai, the best of the four \formulas: 
the \textit{acc@10} and the median wasted effort improve only marginally, and $R_{wef}$ degrades.

\begin{table}[t!]
\caption{RQ1: The effectiveness of GP evolved \formulas using Ochiai, Barinel, Tarantula, and DStar. \label{tab:RQ1-GP}\centering}
\vspace{-0.5em}
\centering
\scalebox{0.85}{
\begin{tabular}{l|r|rrrr|rr}
\toprule
\textbf{Project} & \textbf{Total} & \multicolumn{4}{c|}{\textbf{acc}} & \multicolumn{2}{c}{\textbf{wef (R$_{wef}$)}} \\
& & @1 & @3 & @5 & @10 & mean & med \\
\midrule

Hbase & 8 & 1 & 4 & 5 & 5 & 13.12 (16) & 2.5 (5) \\
Ignite & 14 & 0 & 3 & 3 & 5 & 214.93 (21) & 20.0 (4) \\
Pulsar & 10 & 3 & 5 & 6 & 9 & 9.20 (23) & 3.0 (9) \\
Alluxio & 3 & 0 & 0 & 0 & 1 & 101.67 (65) & 86.0 (83) \\
Neo4j & 3 & 1 & 2 & 2 & 2 & 23.33 (43) & 1.0 (18) \\

\midrule
Total & 38 & 5 & 14 & 16 & 22 & 94.24 (26) & 6.5 (8) \\
Percentage (\%) & 100 & \textbf{13} & \textbf{37} & 42 & \textbf{58} & - & - \\
\bottomrule
\end{tabular}
}
\vspace{-2.2em}
\end{table}

\begin{figure}[ht]
\vspace{-2.2em}
\centering
\includegraphics[width=0.35\textwidth, trim=40mm 0mm 40mm 30mm, clip]{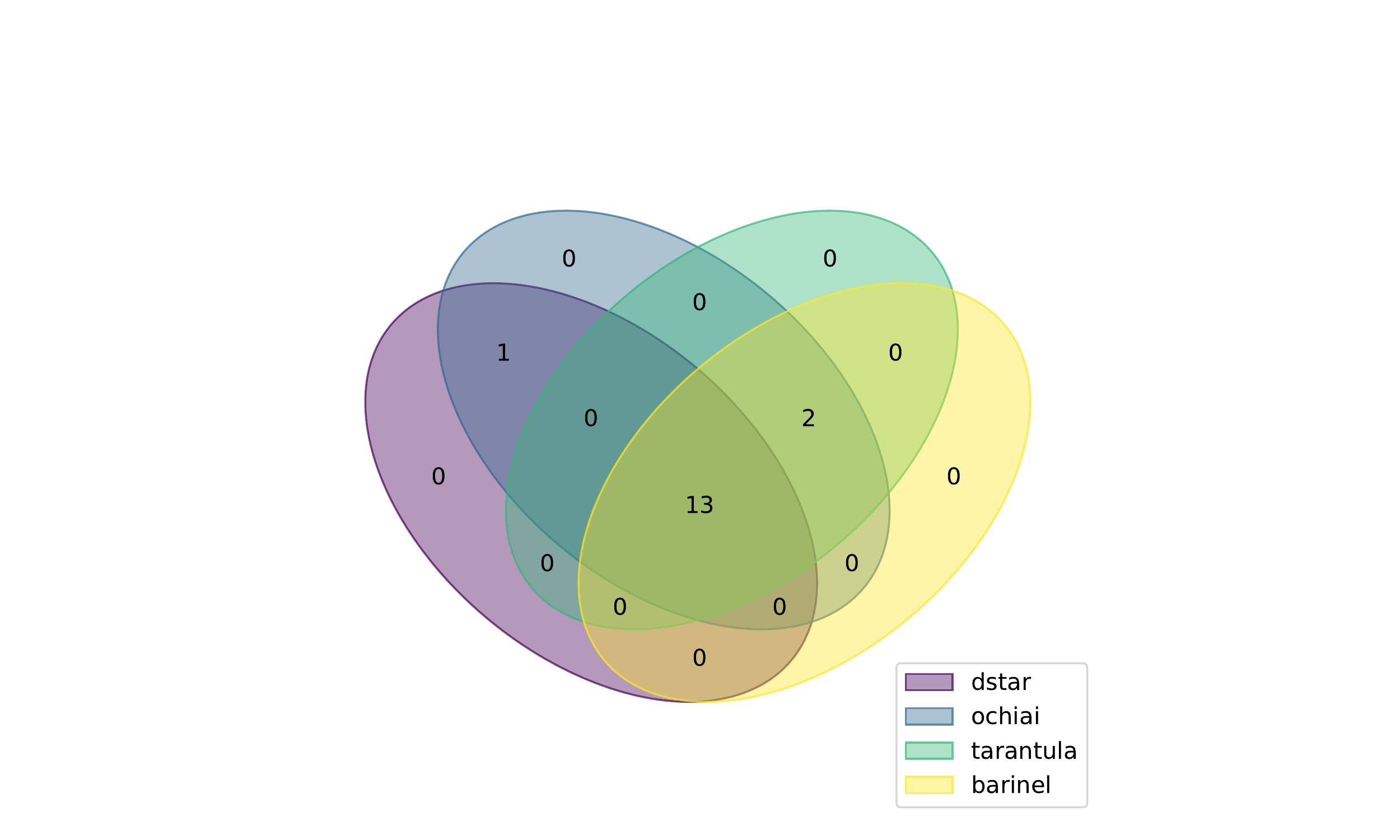}
\vspace{-0.8em}
\caption{Venn-diagram of flaky classes ranked in the top 5 by the four SBFL \formulas.} 
\label{fig:sbfl_venn}
\vspace{-0.6em}
\end{figure}

To understand these observations, we inspect the intersection between the sets of classes ranked in the top 5 by these four SBFL \formulas. Figure~\ref{fig:sbfl_venn} presents this intersection in a Venn diagram.
Out of 14,16,15,15 flaky classes ranked within the top 5 by Dstar, Ochiai, Tarantula, and Barinel, 13 of them are the same flaky classes. There are two additional classes that are ranked in the top 5 by all except Dstar and one extra class by only Ochiai and Dstar.
Overall, the diagram demonstrates that there are large overlaps between the results of these four SBFL \formulas. 
Thus, we can conclude that the GP-evolved formula did not lead to substantial improvements because there was no space for improvement as all four input \formulas provided similar signals. 
This conclusion brings out the need for introducing external signals from other code and change metrics, which will be discussed in the following research question.


\begin{tcolorbox}
Using SBFL, we were able to localise flaky classes by inspecting only 21-24\% (6-7\%) of classes covered by flaky tests on average (median). With Ochiai, flaky classes are ranked at the top and in the top 10 for 16\% and 55\% of total flaky commits. 
\end{tcolorbox}

\subsection{RQ2 - Code and change metrics}

Table~\ref{tab:RQ2} shows the evaluation results for GP-evolved models using SBFL scores with change and code metrics.
The table shows that the addition of signals from change and size metrics leads to an improvement in the identification of flaky classes.
In particular, by adding change metrics, the percentage of classes ranked at the top reaches 24\%. 
This percentage is much higher than the maximum percentage achieved with SBFL alone, which is 16\% with Ochiai.
On the contrary, we do not observe any significant improvements in the number of flaky tests ranked in the top 10 or top 5. 
Combined, these results imply that these change and size metrics can give additional signals that break ties between the classes located near the top, allowing developers to identify the exact cause of flakiness more precisely.
The comparison with the results of GP with only SBFL \formulas in Table~\ref{tab:RQ1-GP} further supports the usefulness of change and size metrics.
Specifically, by adding change and size metrics, the percentage of flaky classes ranked at the top ($acc@1$) goes from 13\% to 24\% and 18\%, respectively. In addition, average $R_{wef}$ improves 5\% with change metrics and 4\% with size metrics. 

With regard to flakiness metrics, their combination with SBFL scores does not lead to any notable improvements in the ranking of classes at the top. 
The percentage of classes at the top is 11\% and the percentage of classes in the top 10 is 53\%.
One possible explanation for this is that our flakiness metrics are derived from a flakiness taxonomy that focuses on the test instead of the CUT.
Hence, using metrics derived from such categories may not be helpful in the identification of CUT components that are responsible for flakiness.
To alleviate this, future studies should consider categories and metrics that are derived from the CUT, and existing flakiness taxonomies should be updated accordingly.

\begin{table*}[ht]
\caption{RQ2: The contribution of flakiness, change, and size metrics to the identification of flaky classes. 
\label{tab:RQ2}\centering}
\centering
\scalebox{0.85}{
\begin{tabular}{c|rrrr|rr|rrrr|rr|rrrr|rr}
\toprule
& \multicolumn{6}{c|}{\textbf{SBFL \& flakiness}} & \multicolumn{6}{c|}{\textbf{SBFL \& change}} & \multicolumn{6}{c}{\textbf{SBFL \& size}}  \\
\textbf{Proj. (\#)} & \multicolumn{4}{c}{acc} & \multicolumn{2}{c|}{wef (R$_{wef}$)} & \multicolumn{4}{c}{acc} & \multicolumn{2}{c|}{wef (R$_{wef}$)} & \multicolumn{4}{c}{acc} & \multicolumn{2}{c}{wef (R$_{wef}$)}  \\
& @1 & @3 & @5 & @10 & mean & med & @1 & @3 & @5 & @10 & mean & med & @1 & @3 & @5 & @10 & mean & med \\
\midrule

Hbase (8) & 1 & 4 & 5 & 5 & 11.9 (12) & 3 (4) & 2 & 4 & 4 & 5 & 16.9 (13) & 4 (4) & 2 & 4 & 5 & 5 & 11.4 (12) & 3 (3) \\
Ignite (14) & 0 & 2 & 2 & 4 & 230.9 (26) & 63 (4) & 2 & 4 & 4 & 4 & 222.3 (24) & 18 (4) & 1 & 3 & 3 & 5 & 220.1 (24) & 43 (4) \\
Pulsar (10) & 2 & 5 & 6 & 8 & 10.2 (15) & 3 (8) & 3 & 5 & 7 & 9 & 8.0 (12) & 2 (5) & 2 & 5 & 7 & 9 & 6.9 (13) & 2 (6)  \\
Alluxio (3) & 0 & 0 & 1 & 1 & 97.7 (51) & 73 (65) & 0 & 0 & 1 & 1 & 75.7 (49) & 94 (39) & 0 & 0 & 1 & 1 & 90.7 (49) & 77 (58) \\
Neo4j (3) & 1 & 2 & 2 & 2 & 19.3 (42) & 1 (18) & 2 & 2 & 2 & 2 & 6.7 (37) & 0 (9) & 2 & 2 & 2 & 2 & 23.0 (40) & 0 (10) \\

\midrule
Total  (38) & 4 & 13 & 16 & 20 & 99.5 (24) & 8 (8) & 9 & 15 & 18 & 21 & 94.1 (21) & 5 (6) & 7 & 14 & 18 & 22 & 94.3 (22) & 5 (7) \\
Percentage (\%) & \textbf{11} & 34 & 42 & \textbf{53} & - & - & \textbf{24} & 39 & 47 & 55 & - & - & \textbf{18} & 37 & 47 & 58 & - & - \\

\bottomrule
\end{tabular}
}
\vspace{-4mm}
\end{table*}

To further investigate the impact of change and size metrics on the identification performance, we analyse the involvement of each metric in our GP-evolved \formulas. 
Table~\ref{tab:RQ2-distr} shows the frequency of change and size metrics in the GP evolved \formulas generated under the configuration of using SBFL and change metrics (\ie SBFL \& Change) and the configuration of using SBFL and size metrics (\ie SBFL \& Size). 
As shown in this table, both change and size metrics are frequently involved in the final \formulas, confirming that our observed improvement did not come only from using GP.
Based on these results, we posit that change and size metrics can contribute positively to the identification of flaky classes. 



\begin{table}[ht]
\caption{Frequency of metrics in GP-evolved \formulas (from 0 to 1). `Changes' and `Dev' denote \textit{`Unique Changes'} and \textit{`Developers'}, respectively. The column `SBFL' contains the average frequency of the four SBFL metrics. 
\label{tab:RQ2-distr}}
\centering
\scalebox{0.85}{
\begin{tabular}{l|r|rrr|rrr}
\toprule
 & SBFL & Changes & Dev & Age & LOC & DOI & CC \\
\midrule
SBFL \& Change & 0.45 & 0.50 & 0.37 & 0.53  & - & - & - \\
SBFL \& Size & 0.50 & - & - & - & 0.71 & 0.37 & 0.73 \\
\bottomrule
\end{tabular}
}
\end{table}

\begin{tcolorbox}
The augmentation of Spectrum-Based Fault Localisation with change or size metrics lets more flaky classes be ranked near the top; by adding change metrics, we can rank 24\% flaky classes at the top. In contrast, metrics specific to flakiness categories do not provide any beneficial signals to the identification approach. 
\end{tcolorbox}

\subsection{RQ3 - Ensemble method}

Table~\ref{tab:RQ3-voting} presents the evaluation results for the voting method with 60 GP-evolved models, half from using SBFL and change metrics and the other half from using SBFL and size metrics.
We decided to exclude the models that build on flakiness metrics since their usage did not improve the performance any further.
As explained in Section~\ref{sec:rq3_ensemble_method}, there can be a case where none of the participating models succeeds to vote for the true candidate since individual models vote only for those ranked within the top n. For this case, we report the median of all rankings of the models as an alternative. 

The results show that the voting step further improves the ranking results.
The most notable improvement is the accuracy at the top 3, which reaches 47\%. Although the improvements in the other accuracy metrics are not as noticeable as what we have seen in the accuracy at the top 3, there are constant improvements over the results without voting. 
The average of wasted effort remains almost the same while the median improves from the voting, dropping to 3.5. These results imply that the voting allows those near the top to shift further to higher ranks based on the agreement among the models that exploit and capture different features of flaky classes.
Nonetheless, the constant improvements in $R_{we}$, both per project and in total, suggest that through the voting, we can rank flaky classes further near to the top; for example, in Alluxio, where $R_{wef}$ is always near 50, average $R_{wef}$ reduces to 22 and its median to 10. 
These results imply that voting can leverage the complementarity between different models, further improving the localisation of flakiness.


\begin{table}[ht]
\caption{RQ3: The effectiveness of the voting between 60 different GP-evolved models, 30 from SBFL with change metrics, and 30 from using SBFL with size metrics. `Perc' denotes Percentage \label{tab:RQ3-voting}\centering}
\centering
\scalebox{0.9}{
\begin{tabular}{l|r|rrrr|rr}
\toprule
\textbf{Project} & \textbf{Total} & \multicolumn{4}{c|}{\textbf{acc}} & \multicolumn{2}{c}{\textbf{wef (R$_{wef}$)}} \\
& & @1 & @3 & @5 & @10 & mean & med \\
\midrule

Hbase & 8 & 3 & 5 & 6 & 6 & 9.62 (12) & 1.5 (2) \\
Ignite & 14 & 2 & 4 & 4 & 4 & 228.61 (24) & 17.5 (4) \\
Pulsar & 10 & 3 & 6 & 7 & 9 & 7.30 (12) & 2.0 (5) \\
Alluxio & 3 & 1 & 1 & 1 & 2 & 61.83 (22) & 9.0 (10) \\
Neo4j & 3 & 1 & 2 & 2 & 2 & 19.67 (42) & 1.0 (18) \\

\midrule
Total & 38 & 10 & 18 & 20 & 23 & 94.61 (19) & 3.5 (5) \\
Perc (\%) & 100 & 26 & \textbf{47} & 53 & 61 & - & - \\

\bottomrule
\end{tabular}
}
\vspace{-4mm}
\end{table}


\begin{tcolorbox}
A voting between models based on SBFL, change, and size metrics, provides the best ranking for flaky classes.
47\% of flaky classes are ranked in the top 3 and 26\% of them are ranked at the top. The average $R_{wef}$ further reduces to 19, highlighting the practical usefulness of our approach.  
\end{tcolorbox}

\subsection{RQ4 - Flakiness categories}
Table~\ref{tab:RQ4} presents the performances of the voting method on the different flakiness categories encountered in our dataset.
The \textit{``Ambiguous"} category represents cases where the flaky tests could not be assigned to any of the known flakiness categories.
First, we observe that the most common categories are Concurrency and Asynchronous Waits.
This aligns with observations from previous studies~\cite{Luo2014,Lam2020a,Eck2019} and confirms that the taxonomy adopted for our metrics is adequate for our distribution.
Furthermore, we observe a discrepancy between the performances in different categories.
Classes responsible for Async Waits are well identified with 80\% of the classes in the top 10, and 30\% of them at the top.
Classes responsible for Concurrency also show good performances with 50\% of them in the top 10, and 38\% of them at the top; the average $R_{wef}$ is below ten, eight precisely, meaning we can locate flaky classes by inspecting less than 10\% of the total number of the classes covered by flaky tests. 

Categories such as Time and I/O show much lower performances, with 33\% and 0\% of flaky classes in the top 10, respectively.
Nevertheless, given the low number of instances for these categories, it is hard to discuss or generalise their results.
With only two instances, the category Unordered Collections shows curious results as one class is ranked second and the other one ranked 663. 
To understand the reasons behind the bad ranking, we manually inspected this case\footnote{\url{https://github.com/apache/ignite/commit/188e4d52c2}}.  
We found that the concerned test, \texttt{testUnstableTopology}, was executed twice due to a retry mechanism. Both executions led to failure, but interestingly, we found that the two failures have different causes. One of them is due to a lack of context initialisation and is likely to be the reason behind flakiness. 
As the two failure causes are different, the coverage is also different in them.
Specifically, one of the failures  did not cover the flaky class, and as the coverage of this failure was leveraged in the SBFL, the flaky class was not considered suspicious.
We discuss other reasons responsible for poor ranking in Section~\ref{sec:discussion}.

\begin{table}[ht]
\caption{RQ4: The effectiveness per flakiness category \label{tab:RQ4}\centering}
\centering
\scalebox{0.8}{
\begin{tabular}{l|cccc|cc}
\toprule
\textbf{Flakiness} & \multicolumn{4}{c|}{\textbf{acc}} & \multicolumn{2}{c}{\textbf{wef (R$_{wef}$)}} \\
\textbf{Category} & @1 & @3 & @5 & @10 & mean & med \\
\midrule

Concurrency (16) & 6 (\textbf{38}) & 7 (44) & 7(44) & 8 (\textbf{50})& 147.53 (27) & 9.5 (9) \\
Async wait (10) & 3 (\textbf{30}) & 6 (60) & 8 (80) & 8 (\textbf{80})& 21.05 (8) & 1.5 (3) \\ 

Ambiguous (4) & 1 (25) & 2 (50) & 2 (50) & 3 (75)& 18.88 (5) & 3.5 (5) \\
Time (3) & 0 (0) & 0 (0) & 0 (0) & 1 (\textbf{33})& 88.33 (16) & 14.0 (10) \\
Network (2) & 0 (0) & 2 (100) & 2 (100) & 2 (100)& 1.00 (10) & 1.0 (10) \\
Unordered  &  &  &  &  & &  \\
collections (2) & 0 (0) & 1 (50) & 1(50) & 1 (50) & 331.5 (33) & 331.5 (33) \\
I/O (1) & 0 (0) & 0 (0) & 0(0) & 0 (\textbf{0})& 12.50 (3) & 12.5 (3) \\
Random (1) & 0 (0) & 1 (100) & 1 (100) & 1 (100)& 2.00 (75) & 2.0 (75) \\


\midrule
Total (39\tablefootnote{One flaky class belongs to two categories: Network and Unordered Collections.}) & 10 & 18 & 20 & 23 & 94.47 (19) & 3.5 (5) \\
Perc (\%) & 26 & 47 & 53 & 61 & - & - \\

\bottomrule
\end{tabular}
}
\vspace{-4mm}
\end{table}

\begin{tcolorbox}
The most prominent flakiness categories, Concurrency and Asynchronous Waits, are identified effectively, with 38\%
and 30\% of their flaky classes ranked at the top, respectively.
In the Concurrency category, flaky classes are identified by examining 8\% of classes covered by flaky tests on average. 
\end{tcolorbox}

\section{Discussion}
\label{sec:discussion}
In this section, we discuss our results in light of the existing literature on test flakiness and fault localisation.
Our approach uses existing fault localisation techniques to identify flaky classes in the CUT.
While we leverage various data sources, the main strength of our approach comes from adopting existing SBFL techniques, as explained in RQ2. The effectiveness of other data, such as change metrics, is limited in providing additional signals that break ties between the classes already ranked near the top. Hence, the performance of our approach largely depends on the applicability of SBFL to our flaky class identification problem.  
The flaky class identification problem and traditional fault localisation problems are similar in the way they are debugged (\ie from the reproduction and cause identification to the fix). As described in~\ref{sub:rq1_effectiveness}, this resemblance allows us to redefine SBFL techniques to identify flaky classes instead of faulty ones. Nevertheless, there is one significant difference between them: the characteristics of a test suite. 
Many fault localisation studies assume a test to cover a single functionality, and the subjects they studied often follow this assumption~\cite{Lou:2021:fse,Li:2019:issta,Wen:2019:tse}.
In contrast, we did not consider such an assumption for test subject selection to reflect a realistic scenario of flaky test failure. 
This difference may restrict the applicability of existing fault localisation techniques to the flaky class identification problem, especially test coverage-based techniques, such as SBFL.
Indeed, although we identified 26\% and 61\% of flaky classes at the top and within the top 10, we failed to reach the performance reported in prior work on fault localisation~\cite{Wong:2016:tse}. Hence, we investigate 
the diagnosability of the test suite of our subjects using the Density, Diversity, and Uniqueness (DDU) metric~\cite{perez:2017:icse}. 

DDU diagnoses the adequacy of SBFL for a software system by considering three properties of its test suite: Density, Diversity, and Uniqueness. Each property covers a distinct feature of a test suite, and DDU is computed as the multiplication of these three properties. 
Density evaluates how frequently a code entity, in our case a class, is covered by tests. Diversity is about whether tests cover code entities in a diverse fashion. Lastly, uniqueness guarantees that different code entities are covered by different sets of tests. 
All these three components of DDU have values between 0 and 1. The higher the DDU is, the more adequate the test suite is for SBFL. 

Table~\ref{tab:DDU} presents DDU values for the five projects analysed in this study. While all five projects generally have high diversity values (\ie all above 0.9), they have relatively low uniqueness and density values, which results in low DDU scores. Among the five projects, Pulsar has the highest DDU score of 0.289, followed by Neo4j, Alluxio, Ignite, and Hbase. Since both Neo4j and Alluxio have only three flaky classes, which might be too small to discuss the identification results, we will skip these two for the following discussion.
Among the remaining three projects, all our flaky class identification methods, ranging from pure SBFL to voting, perform the best on Pulsar, the one with the highest DDU score, in \textit{acc@n} and \textit{wef}. For instance, even the pure SBFL approach that often performs the worst successfully localised nine out of ten flaky classes of Pulsar within the top 10 and more than half within the top five. The same trend was observed in both GP and voting-based methods.
Compared to HBase, while Ignite has a slightly higher average for the DDU score, it has a far lower Uniqueness score (\ie 0.188 for Ignite and 0.413 for HBase). Uniqueness evaluates whether a code entity is distinguishable; we assume that the flaky classes have different coverage than non-flaky classes. Thus, we suspect that Ignite having a lower Uniqueness is why our methods were not as effective on Ignite as on HBase: we have the worst results on Ignite in both absolute (\ie \textit{acc@n} and \textit{wef}) and relative effort (\ie $R_{wef}$). 

Based on these results, we argue that while our outcome may not be as good as those reported by prior fault localisation studies\cite{Yoo:2017ss,sohn-TSE}, that is mainly due to the inherently low diagnosability of a test suite (\eg covering too many classes in the same fashion). This test-suite adequacy issue commonly exists in the fault localisation field\cite{Sohn2021ea} and is not limited to flaky class identification. Hence, we posit that the performance of our approach can improve along with the advances in fault localisation techniques. 

\begin{table*}[ht]
\caption{DDU metrics for the analysed test suites. \label{tab:DDU}}
\centering
\scalebox{0.9}{
\begin{tabular}{l|ccc|ccc|ccc|ccc}
\toprule
\textbf{Project} & \multicolumn{3}{c|}{\textbf{Density}} & \multicolumn{3}{c|}{\textbf{Diversity}} & \multicolumn{3}{c|}{\textbf{Uniqueness}} & \multicolumn{3}{c}{\textbf{DDU}}\\
& min & max & mean & min & max & mean & min & max & mean & min & max & mean \\
\midrule

Hbase & 0.049 & 0.477 &  0.248 & 0.995 & 0.999 & 0.997 & 0.188 & 0.553 & 0.413 & 0.021 & 0.116 & 0.091 \\
Ignite & 0.368 & 0.993 & 0.736 & 0.918 & 1.000 & 0.979 & 0.045 & 0.486 & 0.188 & 0.034 & 0.466 & 0.132 \\
Pulsar & 0.029 & 0.998 & 0.491 & 0.984 & 0.998 & 0.994 & 0.520 & 0.786 & 0.609  & 0.019 & 0.518 & 0.289\\
Alluxio & 0.414 & 0.833 & 0.591 & 0.958 & 0.996 & 0.982 & 0.226 & 0.615 & 0.362 & 0.101 & 0.322 & 0.201  \\
Neo4j & 0.127 & 0.739 & 0.515 & 0.894 & 0.993 & 0.931 & 0.268 & 0.791 & 0.585 & 0.088 & 0.522 & 0.258 \\
\bottomrule
\end{tabular}
}
\vspace{-4mm}
\end{table*}


In an attempt to shed light on the 15 cases where the class was ranked outside the top 10 by our voting approach, we extended our inspection to reason about such performances. We observed that flaky tests covering a high number of classes are more likely to result in low performances. 
For example, the flaky test \texttt{shutdownDatabaseDuringIndexPopulations} in Neo4j covers 480 classes and its flaky class was ranked 59 by our voting approach whereas the other flaky tests in Neo4j (having their corresponding flaky classes ranked 1 and 2) cover fewer than 10 classes. When we inspect the DDU score of the specific commit that contains this test, it has a relatively low DDU score compared to the other two commits. 
Additionally, most of the mis-ranked classes are found in the Ignite project (10/15), whose DDU score is the second-lowest, and its tests cover on average 492 classes. Due to this consequent number of covered classes, we suspect these tests to be of a higher level, \ie integration or end-to-end tests. 
This aligns with studies highlighting the prevalence of flakiness in integration and system tests~\cite{Kowalczyk2020,Herzig2015}.
Still, our approach does not systematically fail to identify flaky classes covered by higher-level tests as nine of them (flaky test covering more than 100 classes) are listed in the top 10.
\section{Threats to Validity}\label{sec:threats}

\paragraph{External validity}
The main threat to the external validity of this study is the dataset size.
To ensure the generalisability of our results, it would have been preferable to include more flaky tests in our experiments. 
Nonetheless, the datasets of flaky tests are generally limited in size due to the elusiveness of flakiness~\cite{habchi2021mutinject,Haben2021,alshammari2021flakeflagger}.
Moreover, as explained in Section~\ref{sec:data_collection}, the requirements of this study limited the set of candidates considerably.
For a commit to be eligible in our study, it needs to have atomic changes fixing flakiness in the CUT.
However, only 24\% of flaky tests actually stem from the CUT, which limits the size of potential subjects~\cite{Luo2014}.
Besides, the creation of our dataset required a substantial amount of manual work to identify suitable commits and perform necessary changes to retrieve coverage matrices.
For instance, for each commit, we had to modify the build script to match \textsc{Gzoltar} requirements, \ie find the test executor version that matches both the program under test and the plugin.
We iteratively removed non-essential listeners and other plugins that could interfere with the instrumentation.
Moreover, we had to find and adapt the execution environment to match the program under test and the testing environment.
Finally, compared to the works of Lam~\etal~\cite{Lam2019RootCausing} and Zitfci and Cavalcanti~\cite{De-Flake}, which were conducted on proprietary software, this study is the first to leverage open-source software to localise flakiness root causes.
Thus, our dataset and ground truth can be valuable for future studies on flakiness debugging.


\paragraph{Internal validity}
One potential threat to our internal validity is our definition of flakiness root causes within the CUT, \ie flaky classes.
We rely on flakiness-fixing commits to identify classes that are responsible for flakiness.
However, we cannot be certain that (i) the flakiness fix is effective, and (ii) the modified class is the one responsible for flakiness.
Indeed, a study by Lam~\etal~\cite{Lam2020a} showed that developers may wrongly claim that their commits fix flaky tests before realising that the fix is ineffective.
Additionally, there are no guarantees that the classes included in the fix are the ones responsible for flakiness.
Nonetheless, if the class was part of the proclaimed fix, this means that the developers found it, at least, relevant. Hence, its identification by our approach is still helpful for developers willing to understand, debug, and fix flaky tests.

\paragraph{Construct validity} 
One potential threat to our construct validity is our measurement of the coverage for flaky tests.
A flaky test can pass and fail for the same version of the program, but in practice, it may be extremely difficult to reproduce both the pass and failure~\cite{alshammari2021flakeflagger,Lam2020}.
Hence, a test can be observed as flaky by the project developers and therefore fixed, yet we are unable to reproduce the pass and failure in our experiments even with a large number of reruns~\cite{alshammari2021flakeflagger}.
For this reason, we focused on the available status, \ie pass or failure, and retrieve its coverage.
It is possible that including the coverage of both the pass and failure from the flaky tests might lead to different results with spectrum-based fault localisation.
Thus, we encourage future studies to investigate this direction.
Another possible threat is whether the evaluation results of our approach truly support what we claim. 
We use two absolute metrics, $acc@n$ and $wef$, that can reflect the realistic debugging effort of developers, following the suggestion from Parnin and Orso~\cite{parnin}, and one relative metric, $R_{wef}$, to compare with the baseline of inspecting classes covered by flaky tests. 
\section{Related Work}
\label{sec:related_works}
\paragraph{Flakiness root causes}
Several empirical studies highlighted the diversity of flakiness root causes. Luo~\etal~\cite{Luo2014} were the first to characterise the root causes of flaky tests. 
They analysed 201 flakiness-related commits from 51 open-source projects and showed that the mismanagement of asynchronous calls and concurrency are the most common causes of flaky tests. 
Later studies replicated the work of Luo~\etal, showing that other flakiness root causes can be more relevant in different application domains. 
Thorve~\etal~\cite{Thorve2018} analysed 77 flakiness-related commits in 29 open-source Android applications and found that 22\% of these commits have flakiness caused by external factors like hardware, operating system version, and third-party libraries. 
Eck~\etal~\cite{Eck2019} surveyed 21 Mozilla developers, asking them to classify 200 flaky tests in terms of root causes and fixing efforts. The survey results highlighted four new categories of flakiness: restrictive ranges, test case timeout, test suite timeout, and platform dependency. 

\paragraph{Flakiness root cause analysis}
The main contribution to flakiness root cause localisation was proposed by Lam~\etal~\cite{Lam2019RootCausing}. 
They introduced a framework that helps developers to localise the root causes of their flaky tests. This framework uses an instrumentation tool to log the runtime properties of the test execution. Then it reruns the tests 100 times to produce logs for a passing and a failing execution. To analyse these logs and localise the root cause, they propose RootFinder, a tool that compares the logs of passing and failing executions to identify methods that can be responsible for flakiness. RootFinder relies on a predefined set of non-deterministic method calls and does not explore calls of unknown methods. Hence, it can only detect flaky tests that arise from method calls that the developer is already suspecting. 
Zitfci and Cavalcanti~\cite{De-Flake} presented Flakiness Debugger, a tool that compares the code coverage of passing and failing executions to localise the flakiness root cause. They ran their tool on 83 flaky tests and presented the localised root cause to two developers asking them for their evaluation. On average the developers found that in 48\% of the cases, flakiness was due to the exact statements spotted by Flakiness Debugger. Moreover, only 18\% of the outputs were considered inconclusive, hard to understand, or not useful.
Both RootFinder and Flakiness Debugger relied on differences between passing and failing executions of flaky tests to localise flakiness in the CUT. In this study, we explore a new direction by analysing the differences between flaky and stable tests.
  
Morán~\cite{flakyloc} presented FlakyLoc, a tool for localising the root causes of flakiness in web applications. The tool reruns web tests while varying environmental factors (network, memory, CPU, browser type, operating system, and screen resolution) and records test results. Then, it uses ranking metrics (Ochiai and Tarantula \cite{sbfl-evaluation,tarantula}) to identify the environmental factor and value that are responsible for the flaky failure. The tool was only evaluated on one test case and it detected that the failure was caused by low screen resolution. 
In this paper, we do not focus on any specific flakiness category and our analysis is based on the test coverage instead of environmental factors.

\paragraph{Fault localisation}
Fault localisation was introduced to ease the developers' burden of debugging by automatically identifying the root cause of a program failure~\cite{Catal2011fc,Wong:2016:tse}. 
Since its appearance, various fault localisation techniques that utilise various data, from dynamic to static ones, have been actively proposed~\cite{Wen:2012icse,Papadakis:2015sf,Li:2019:issta,wong2016survey}. 
Spectrum Based Fault Localisation (SBFL), a lightweight coverage-based technique, was under the spotlight for many years. SBFL takes a test coverage matrix as input and computes the likelihood of containing a fault for individual code entities using a risk evaluation formula. 
The simplicity and effectiveness of SBFL attract many researchers into this field~\cite{Wong:2007rt,Xie:2013uq,Xie:2013kx,Yoo:2017ss}.
Papadakis and Traon proposed Mutation Based Fault Localisation (MBFL) techniques that leverage the coupling between real faults (\ie complex faults) and the mutants (\ie simple faults) to localise faults in code~\cite{Papadakis:2015sf}.  
Information Retrieval-based Fault Localisation (IRFL) approaches the problem differently, utilising static data sources, such as bug reports, instead of dynamic ones, such as test coverage~\cite{wong2016survey}. 
Recently, Li \etal proposed to combine various fault localisation techniques using a deep learning model~\cite{Li:2019:issta}. Their approach called DeepFL successfully outperforms all the other FL techniques that it considers. 
The current trend of fault localisation is moving to either use a deep neural network to train a model~\cite{Lou:2021:fse,Li0N21a} or include humans to bring additional signals~\cite{Li:2018:icse}. In which direction it heads, the main framework of fault localisation remains the same and test coverage remains to be an effective source of information. 



\section{Conclusion}
\label{sec:conclusion}
We presented the first empirical evaluation of SBFL as a potential approach for identifying flaky classes. We investigated three approaches: pure SBFL, SBFL augmented with change and code metrics, and an ensemble of them. 
We evaluated these approaches on five open-source Java projects. Our results show that SBFL-based approaches can identify flaky classes relatively well, especially with code and change metrics, suggesting that code components responsible for flakiness exhibit similar properties with faults. This finding highlights the potential of existing fault localisation techniques for flakiness identification. At the same time, the results show that flaky tests can have unique failure causes that may mislead any coverage-based root cause analysis, stressing the need to consider these flakiness-specific causes in future studies.
%
%
%

Our study forms the first step towards flakiness localisation. We believe that there is a lot of room for improvement and encourage future studies to explore additional techniques, fault prediction metrics, and devise techniques that can further improve and support flakiness localisation.

\bibliographystyle{IEEEtran}
\bibliography{references} 
\end{document}